# A Comparative Study of Web Services Composition Networks

Chantal Cherifi
SPE Laboratory
Corsica University
Corte, France
chantalbonner@gmail.com

Jean-François Santucci
SPE Laboratory
Corsica University
Corte, France
santucci@univ-corse.fr

*Abstract*—Web services growth makes the composition process a hard task to solve. This numerous interacting elements can be adequately represented by a network. Discovery and composition can benefit from the knowledge of the network structure. In this paper, we investigate the topological properties of two models of syntactic and semantic Web services composition networks: dependency and interaction. Results show that they share a similar organization characterized by the small-world property, a heavy-tailed degree distribution and a low transitivity value. Furthermore, the networks are disassortative.

*Keywords; Web services; composition, Web services networks; complex networks*

## I. INTRODUCTION

Web services are web-based software applications designed to be published, discovered and invoked for remote use. Those modular applications can be programmatically loosely coupled through the Web to form more complex ones. Two of the most popular problems in Web services technology are discovery and composition. Discovery is the process of locating providers advertising Web services that can satisfy a service request. Composition arises when several Web services are needed to fulfill a request. The basic architecture of WSDL, SOAP and UDDI is insufficient to realize truly automatic Web services discovery and composition. To overcome this drawback semantic Web service descriptions such as WSDL-S, SAWSDL, OWL-S and WSMO have been proposed. Despite all these efforts, Web services discovery and composition are still highly complex tasks. The complexity, in general, comes from different sources. Among them is the scale effect; Web services are numerous on the Web and their number is continuously growing. Another source of complexity is their volatile aspect; providers may change, relocate, or even remove them. Consequently, the Web services space is an evolving structure of a great number of atomic Web services. Knowledge of the Web services overall structure becomes of paramount importance. It is a key for optimizing discovery and composition processes. In the composition context, the Web services space can naturally be represented by a network of interacting atomic Web services.

Despite their great potential, such representations have not widely spread out in the Web services community. Existing research are mainly concerned with Web services network representation for composition mining purposes [2], [2], [3], [4]. To our knowledge, the only attempt to provide a topological landscape of Web services networks is related in [5], [6]. The authors take benefits of their results in order to build a generator of synthetic Web services descriptions and to propose a composition search algorithm. Although this work covers a wide range, it only focuses on syntactically described Web services. Furthermore, in all those previous work, none of them explore the relations between the different network models. Mostly, they concentrate on one type of Web services composition network, sometimes not clearly defined and none of them attempts to compare syntactic and semantic Web services descriptions.

In this work, our motivation is to provide a comparative evaluation of the topological structure of Web services composition networks models. Under the complex network paradigm, we investigate the topological properties of the networks from to entry points: the model (dependency, interaction) and the description (syntactic, semantic). In dependency networks, nodes are Web services parameters and links are dependency relation between parameters. Hence, the connections between nodes represent a production flow. In interaction networks, nodes are operations and links are interaction between two operations. In this case, the connections between nodes represent an information flow. To identify nodes in a dependency network or links in an interaction network, we rely on parameters similarity. We use an equal matching method for syntactic described Web services and an exact matching method for semantic described Web services. To investigate the Web services networks properties, we choose an appropriate collection of publically available Web services descriptions. It contains a large number of real-world Web services descriptions both syntactic and semantic.

This article is organized as follows. In section 2, we recall fundamental elements regarding Web services definition, discovery and composition. In section 3, we present dependency and interaction network models and give the details of the matching methods. We review the network properties used in the network analysis in section 4. A presentation of the experimental methodology and the results are reported in section 6. Finally, we end the article in section 7 by discussing some conclusions and directions for future work.

## II. WEB SERVICES

In this work, we are mainly concerned by two aspects of Web services, namely description and composition. We first give a definition of Web services. Then we tackle the Web services description aspect by presenting the two ways of describing Web services, syntactic and semantic. We finally give a definition of Web services composition with an illustrative example.

## A. Definition

There is no common understanding in the literature of what Web services are. In [7] an interesting and detailed discussion is provided on existing definitions. The most refined is the one given by the W3C. It envisions a Web service as a "software application identified by a URL, whose interfaces and bindings are capable of being defined, described and discovered as XML artefacts. A Web service supports direct interactions with other software agents using XML-based messages exchanged via Internet-based protocols." This definition hints at how Web services should work. It stresses that Web services should be capable of being "defined, described, and discovered". Hence, it is possible to have clients that bind and interact with them. In other words, Web services are components that can be integrated into more complex distributed applications. A common point, on which all the definitions agree, is the fact that a Web service is a distributed application that exports a view of its functionalities. Such a view can be described from an input/output perspective. This is the one retained in our work. Hence, a Web service consists of a set of operations. An operation $i$ represents a specific functionality. It is characterized by one set of input parameters noted $I_i$, and one set of output parameters noted $O_i$. $I_i$ is the required information in order to invoke a Web service operation $i$. $O_i$ is the provided information by the Web service operation $i$. Fig. 1 represents a Web service labeled α with two operations numbered 1 and 2, and the sets of input and output parameters. Note that two level of granularity can be considered. In a "white box" approach the atomic elements are the operations. In this case each operation is described by its input/outputs sets. In a "black box", approach the atomic elements are the Web services. In the following we consider only the former case.

## B. Description

The description of Web services capabilities is essential for Web services discovery, composition and management. Two trends are followed to describe Web services: syntactic and semantic. For our concern, the main difference between a syntactic description and a semantic one takes place at the parameter level.

In the syntactic case, the parameter's associated information is its name and its XML data type. To syntactically describe Web services, Web Service Description Language (WSDL) has been proposed in the context of the W3C. This XML-based language specifies a Web service by defining messages and operations. Each message can consist of one or more parameters. As the parameter name is given by each provider without any consensus, misinterpretations may occur.

Semantic Web services aim at augmenting Web services with rich formal descriptions of their capabilities. This is realized through the use of ontologies. An ontology is defined as a formal and explicit specification of a shared conceptualization [8]. In this case, an ontological concept describes parameters. This formal, explicit specification originating from a shared vocabulary and taxonomy model of a domain is supposed to be unambiguous.

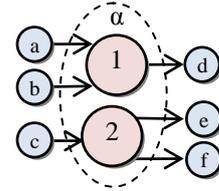

Figure 1. Schematic representation of a Web service α with two operations 1 and 2. $I_1=\{a,b\}$, $O_1=\{d\}$, $I_2=\{c\}$, $O_2=\{e,f\}$. At the Web service level, the set of input parameters of a Web service α is $\{a,b,c\}$ and the set of its output parameters is $\{d,e,f\}$.

In order to define semantically described Web services, two main conceptual approaches have been developed. The first approach aims at semantically annotating existing WSDL descriptions of Web Services. WSDL-Semantic (WSDL-S) [9] and Semantic Annotation for WSDL (SAWSDL) [10] are two lightweight ways of adding semantics to WSDL Web services descriptions. Those W3C specifications allow establishing mapping between existing WSDL elements and ontological concepts. The second approach aims at developing pure semantic Web services. The field includes substantial bodies of work, such as the efforts around Ontology Web Language for Services (OWL-S) [11]. OWL-S is an ontology of Web services. This W3C specification consists of a set of three OWL sub ontologies. Among them, the service profile is designed for advertisement and discovery. It includes general information about the Web service such as its name, its parameters and preconditions and effects, which are mapped to ontological concepts.

## C. Composition

Web service composition addresses the situation when a client request cannot be satisfied by any available atomic Web service. In this case, a composite Web service can fulfill the request. A composite Web service is obtained by combining existing available atomic or even other composite Web services.

Web service composition is twofold. It comprises composition synthesis and composition orchestration. Given a set of available Web services and a client request, the problem of composition synthesis, or simply composition, is concerned with synthesizing a new composite Web service. It thus produces a specification of how to link the available Web services to realize the client request. The problem of orchestration then deals with the coordination of the various Web services and the monitoring of the data flow among them.

The following example illustrates the composition synthesis problem. A user wants to get the publication date of a book, Fig. 2(a). He knows the name of the author and the title of the book. In case where there is a Book Web service which gives a book publication date, if one provides it with an author name and a book title his request may be satisfied by this atomic Web service. If such a Web service does not exist, several other existing Web services can be combined. Assume the two following Book Web services are available, Fig. 2(b): the first one, `AuthorNameBookTitle_ISBN`, provides a book ISBN number against the provision of the author name and the book title; the second one, `ISBN_PubliDate`, provides a book publication date against the provision of the book ISBN number. A new composite Web service can be synthesized that fully satisfy the user request. It is obtained by combining `AuthorNameBookTitle_ISBN` and `ISBN_PubliDate` Web services as follows, Fig. 2(c). First, the user provides the information he knows, that is author name and book title, in order to use the `AuthorNameBookTitle_ISBN` Web service. Second, the information provided by this first Web service is used by `ISBN_PubliDate` Web service which in turn provides the user's final expected information, namely the publication date of the book.

## III. WEB SERVICE NETWORKS

In this section we present the main models of composition networks namely dependency and interaction. Note that more sophisticated interaction network models can be defined such as bipartite graphs and hypergraph based models.

### A. Dependency Network

A dependency network is a directed graph where the nodes are the set of parameters and a link is drawn from an input to an output parameter of any operation. A dependency network expresses the dependency relationship between input and output parameters. In other words, the production of the output parameter depends on the provision of the input one through the invocation of a Web service operation. In such networks, the presence of a link from a node $p_1$ towards another node $p_2$ indicates that at least one operation uses the parameter corresponding to $p_1$ as an input and the parameter corresponding to $p_2$ as an output. The lower left part of Fig. 3 represents the dependency network issued from the Web services α, β and γ depicted on the upper part of the figure.

### B. Interaction Network

An interaction network is a directed graph where the nodes are the set of operations and a link indicates that the output parameters of an operation provide the necessary information to invoke another one. One can consider different levels of interactions depending on the degree of information provided by the invoking operation. The partial interaction corresponds to the case where it provides only a portion of the input parameters to the invoked operation. Full interaction represents the case where all the parameters are provided. In other words, in a full interaction network, a link is drawn from an operation $i$ towards another operation $j$ if and only if for each input parameter in $I_l$, a similar output parameter exists in $O_k$. Partial interaction networks are interesting to deal with optional input values, while full interaction is more suited to the situations where all data inputs are necessary. In the following we will exclusively consider the full interaction mode. The down right side of Fig. 3 corresponds to the full interaction network associated to the Web services presented in the upper part.

### C. Parameters Similarity

A central task in extracting composition networks is to determine if two parameters are similar. Indeed, in a dependency network, parameters similarity is used to determine if some parameters have to be represented by the same node while in an interaction network it is used to decide if a link has to be drawn between two nodes. Furthermore, this task depends on the nature of the considered parameters, syntactic or semantic.

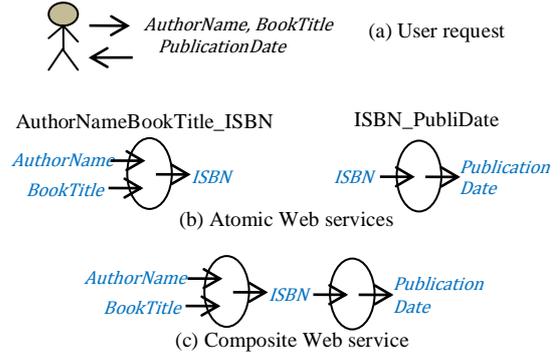

Figure 2. Example of a Web service Composition (c) elaborated form a request (a) and two atomic Web services (b). The ISBN number provided by `AuthorNameBookTitle_ISBN` is used by `ISBN_PubliDate`, to answer the user request.

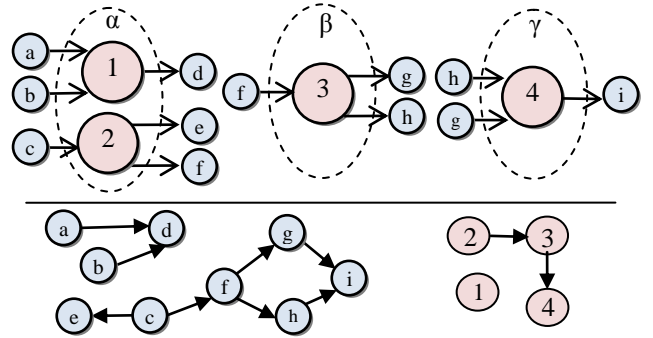

Figure 3. Interaction network of parameters with 9 nodes labeled from a to i (down left) and interaction network of operations with 4 nodes labeled from 1 to 4 (down right) obtained from four operations (top).

For syntactic descriptions, the similarity is assessed on parameters name through a syntactic matching. In [12], we studied different metrics used to build Web service composition networks. We concluded that, when properly used, their influence on the network topology is negligible. Consequently, in this work, we consider only the strict similarity, i.e. two parameters are said to be similar if their names are exactly the same strings. This matching case is called equal.

For semantic descriptions, the similarity is performed on parameters concept. To measure the similarity between two ontological concepts, we use also a strict semantic matching based on the classic exact operator introduced in previous Web services-related works to compare ontological concepts [13]. The exact matching considers two parameters as similar if their concepts match perfectly.

## IV. NETWORK PROPERTIES

We recall the definition of the most useful properties which summarize the essential of a network structure from the complex network perspective.

### A. Small-World

A small-world network is a network in which most nodes are not neighbors of one another, but most nodes can be reached from every other by a small number of hops or steps. This notion is related to the network distance. The distance between two nodes is defined as the number of links in the shortest path connecting them. The small-world property is observed when the average distance of the network is relatively small. The classic procedure to assess this property consists in comparing the average distance measured in some network of interest to the one estimated for an Erdős-Rényi network containing the same numbers of nodes and links. Indeed, this model is known to produce networks exhibiting the small-world property [14].

*B. Scale-Free*

The degree distribution has significant consequences for our understanding of natural and man-made phenomena. It expresses the fraction of vertices in the network with degree *k*. For directed networks there is an out-degree distribution, an in-degree distribution and the joint in-degree and out-degree distribution. Most networks in the real-world, notably the Internet, the World Wide Web and some social networks, have a degree distribution which is very different from the binomial distribution of a random graph. Their degree distribution is highly inhomogeneous. A large majority of nodes have low degree and a small number have high degree. This degree distribution approximately follows a power law. It is known as scale-free degree distribution [14], [15], [16] and corresponding networks are called scale-free networks.

*C. Transitivity*

The transitivity corresponds to the density of triangles in a network. A triangle is a structure of three completely connected nodes. It is measured by the ratio of existing triangles to possible triangles in the considered network [17]. The higher the transitivity is, the more probable it is to observe a link between two nodes which are both connected to a third one. Real-world networks and more particularly social ones generally have a high transitivity value. The transitivity is supposed to be higher than the one of a corresponding Erdős-Rényi network by an order of magnitude corresponding to their number of nodes.

*D. Degree Correlation*

The degree correlation reveals the way nodes are related to their neighbors according to their degree. If high-degree nodes in a network preferentially associate with other high-degree nodes, the network is said to be assortative while if high-degree nodes prefer to attach to low-degree ones, the network is said to be disassortative. Both situations are seen in some networks, as it turns out. The degree correlation values ranges from -1 to 1. Real-world networks usually show a significantly different from zero degree correlation. If it is positive, the network is said to have assortatively mixed degrees. If it is negative, the network is disassortatively mixed [14]. According to Newman [14], social networks tend to be assortatively mixed, while other kinds of networks are generally disassortatively mixed.

## V. EXPERIMENTS

Our main point of interest is to know if the dependency and the interaction models possess the same topological properties. Two main directions can be taken in order to compare the composition networks topological properties. The first one is by considering the description axis: syntactic vs. semantic. The second one is by considering the relationship axis: dependency vs. interaction. Results on the comparative analysis of syntactic and semantic networks are given in [18], [19]. In this work, our main focus is on the relationship. Nevertheless, to provide further insight to this study, we also report the results on both syntactic and semantic networks in order to illustrate the description influence on networks properties.

*A. Set-up*

A few organizations are providing Web services description benchmarks [20], [21], [22], [23]. Among these benchmarks, the SAWSDL-TC collection provided by SemWebCentral [22] is the most appropriate for our experiments. It is big enough to meet our requirements, descriptions contain both syntactic and semantic information and finally, although re-sampled, the collection contains a part of real Web services descriptions. We extracted from SAWSDL-TC a dependency network and an interaction network from the syntactic and the semantic descriptions, using WS-NEXT [24], a tool that we specifically designed for this purpose.

*B. Results and Discussion*

In this section, we first compare the global structure of the networks. Then, focusing on the largest component, we examine their topological properties i.e. small-world, scale-free, transitivity, degree correlation.

*1) Networks global Structure*

All the studied networks share the same global structure. We can identify three parts: a set of isolated nodes, a set of small components and a large component. Fig. 4 and Fig. 5 respectively represent the Syntactic Dependency Network and the Syntactic Interaction Network. Networks are represented without the isolated nodes. Distinct components exist, reflecting the decomposition of the collection into several non-interacting groups of parameters or operations. The presence of a large component is a good property. Indeed, the number of possible dependencies or interactions is high, allowing a large proportion of Web services to participate in a composition.

Characteristics of these three parts of the networks are reported in Table 1. The proportion of isolated nodes is much higher in the interaction networks than in the dependency networks. For example this proportion is 4.2% in the semantic dependency network and 49% in the semantic interaction network. Parameters represented by isolated nodes in a dependency network meet the following criteria. They belong to operations having only input parameters or they belong to operations having only output parameters. As we have few isolated nodes in the networks, this indicates that few parameters have those characteristics. The great majority are shared by operations.

In an interaction network, isolated nodes represent operations that do not interact with others. None of their output parameter can serve as input and none of their input parameter is provided by other operations. Hence, they only can be invoked as atomic operations. In terms of composition, they do not represent any value. In the case of a network mining, the composition search can be performed on components without those components being polluted by unnecessary operations.

Hence, when we compare the two models, we can see that operations that are isolated nodes in an interaction network are found within components in a dependency network.

The number of small components is globally higher in the dependency networks than in the interaction networks. For example, there are 15 small components in the semantic dependency network and 7 small components in semantic interaction network. As shown in Table 1, their size globally ranges between 2 and 30. Nevertheless, there are more very small components in the dependency networks. For example, in the syntactic dependency network, 13 small components on 16 contain only from 2 to 6 nodes. It is interesting to notice that, according to the model itself, the components in each network type do not have the same meaning. In an interaction network, a component necessarily represents a composition. The smallest possible component of two nodes embodies two operations in an interaction relation. This is not the case in a dependency network where a component may represent a single operation. If it contains several operations, they share some parameters but this not imply that there are linked by a composition relation. Hence, the nodes repartition into small components follows the same pattern in both dependency and interaction networks.

In dependency and interaction networks, the largest component contains the majority of the nodes and links. Nevertheless, the largest components of the dependency networks are proportionally smaller with less than 80% of the nodes and less than 90% of the links. In the interaction networks, the largest components roughly contain between 85% and 95% of the nodes. They contain between 95% and 99% of the links. As shown in Table 1, dependency networks are sparser than interaction networks. As the composition search in a less dense network is less costly, dependency networks may be more suitable for composition search. However, this observation must be tempered by the fact that those networks convey the information on possible compositions, but not on the operations entering in those compositions.

2) Networks Properties
   a) Small-world

Both dependency and interaction networks possess the small-world property. Indeed, they exhibit a small average distance. As shown in Table 2, all the composition networks have a smaller distance than the one of the corresponding Erdős-Réyni network. Such results mean that many short cuts exist in the networks. Hence, in a dependency network, one can produce some parameters of interest using a relatively small number of operations. In an interaction network, one can find compositions implementing a request functionality using a relatively small number of operations. Nevertheless, we can observe that within a same description type, the average distance values are always greater for the dependency networks. Furthermore, semantic networks exhibit a lower average distance as compared to their syntactic counterpart. The diameter, although not related to the small world-property, is a good indicator of the largest possible dependency path in the dependency networks. In parallel, it is a good indicator of the largest possible composition in the interaction networks. As shown in Table 2, diameter values are all small regarding the networks size.

TABLE I. NETWORKS COMPONENT ORGANIZATION: NUMBER AND PERCENTAGE OF ISOLATED NODES, NUMBER AND SIZE OF SMALL COMPONENTS, SIZE, NUMBER OF LINKS, PERCENTAGE OF NODES AND LINKS AND DENSITY OF LARGEST COMPONENTS. PERCENTAGES ARE COMPUTED ON TRIMMED NETWORKS

| Properties | Dependency | | Operation | |
|---|---|---|---|---|
| | *Syntactic* | *Semantic* | *Syntactic* | *Semantic* |
| isolated nodes | | | | |
| number | 18 | 15 | 351 | 383 |
| % | 4.67% | 4.2% | 44,71% | 49% |
| small components | | | | |
| number | 17 | 15 | 5 | 7 |
| size | 2-29 | 2-14 | 2-22 | 2-28 |
| largest component | | | | |
| size | 269 | 268 | 395 | 341 |
| number of links | 633 | 621 | 3666 | 3426 |
| nodes % | 73% | 78% | 91% | 85% |
| links % | 86% | 88% | 98% | 98% |
| density | 0.0087 | 0.0086 | 0.0235 | 0.0295 |

TABLE II. SMALL-WORLD PROPERTY OF THE COMPOSITION NETWORKS: AVERAGE DISTANCE AND DIAMETER

| Properties | Dependency | | Interaction | |
|---|---|---|---|---|
| | *Syntactic* | *Semantic* | *Syntactic* | *Semantic* |
| average distance | | | | |
| composition network | 2.75 | 1.97 | 2.19 | 1.87 |
| Erdős-Réyni network | 6.29 | 6.24 | 2.91 | 2.76 |
| diameter | 7 | 5 | 8 | 4 |

Hence, the largest possible composition involves few operations of the whole collection.

   b) Scale-free

An empirical analysis of the dependency networks shows that few parameters have a huge number of links while the majority has only a few. This is characteristic of an inhomogeneous degree distribution. Assuming a power-law distribution, we computed the maximum likelihood exponent of the distribution. Using this value we performed a Kolmogorov Smirnov goodness of fit test to measure the discrepancy between observed values and the values expected under the power-law model. The estimated power-law coefficients and associated p-values are reported in Table 3.

For the dependency networks, we obtain significant p-values for the estimated power-law. Therefore we may reasonably believe that the power-law distribution is an adequate model for the underlying distributions. The p-values are lower for the in-degree distributions and for the out-degree distributions than for the joint-degree distribution. The syntactic out-degree p-value is the only one for which the assumption is questionable.

For all the interaction networks, we obtained almost zero p-values (not mentioned in Table 3) and therefore we conclude that the degree distributions of the interaction networks do not follow a power-law. Nevertheless, we still observe the presence of hubs and authorities corresponding to heavy tailed distributions.

TABLE III. SCALE-FREE PROPERTY OF THE COMPOSITION NETWORKS: EXPONENT Γ AND P-VALUES

| Properties | Dependency | | | | Interaction | |
|---|---|---|---|---|---|---|
| | *Syntactic* | | *Semantic* | | *Syntactic* | *Semantic* |
| | γ | p-value | γ | p-value | γ | γ |
| in-degree distribution | 3.15 | 0.42 | 2.99 | 0.57 | 1.79 | 1.69 |
| out-degree distribution | 2.01 | 0.02 | 3.45 | 0.21 | 2.79 | 2.8 |
| joint-degree distribution | 3.15 | 0.81 | 3.04 | 0.84 | 2.96 | 2.17 |

*c) Transitivity*

For both dependency and interaction networks, transitivity is relatively low. As we can see in Table 4, the transitivity coefficient values of the composition networks are comparable to the transitivity coefficient values of the random networks. Random networks are known to have a very low transitivity coefficient. Hence, we can say that the Web services composition networks do not possess the transitivity property. Indeed, as we can see in Fig. 4 and Fig. 5, nodes are organized hierarchically. This results in a network structure dominated by trees rather than by triangles. This structure favors the apparition of hubs and authorities. Hubs correspond to parameters used as input by many operations in the dependency networks. They correspond to operations used by many others in the interaction networks. Authorities correspond to parameters being outputs of many operations in the dependency networks. They correspond to operations that are used by many others in the interaction networks.

*a) Degree correlation*

Negative values of the degree correlation coefficient reported in Table 5 indicate that nodes are significantly disassortatively mixed. In such a configuration, strongly connected nodes are preferentially linked with lightly connected ones. This is a typical behavior observed in other real-world networks such as information networks, technological networks and biological networks.

TABLE IV. TRANSITIVITY OF THE COMPOSITION NETWORKS

| Properties | Dependency | | Interaction | |
|---|---|---|---|---|
| | *Syntactic* | *Semantic* | *Syntactic* | *Semantic* |
| composition network | 0.039 | 0.031 | 0.032 | 0.022 |
| Erdős-Réyni network | 0.018 | 0.020 | 0.47 | 0.060 |

TABLE V. DEGREE CORRELATION OF THE COMPOSITION NETWORKS

| Properties | Dependency | | Interaction | |
|---|---|---|---|---|
| | *Syntactic* | *Semantic* | *Syntactic* | *Semantic* |
| composition network | -0.21 | -0.22 | -0.45 | -0.51 |

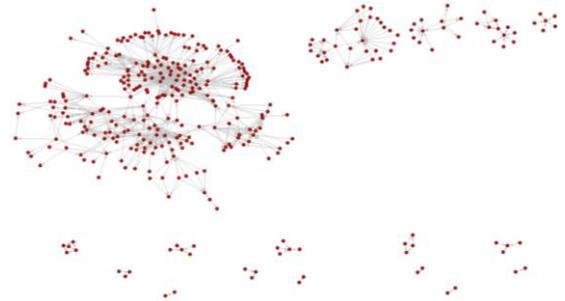

Figure 4. Syntactic dependency network extracted from SAWSDL-TC. Isolated nodes are not represented.

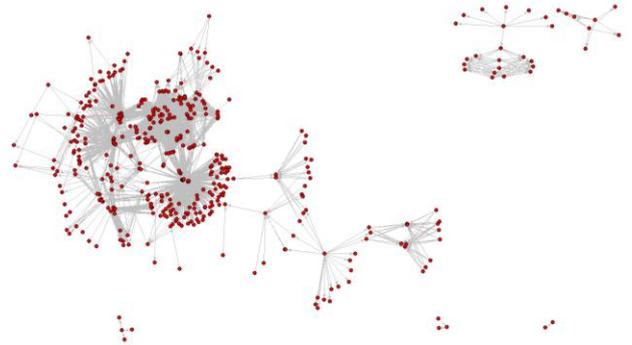

Figure 5. Syntactic interaction network extracted from SAWSDL-TC. Isolated nodes are not represented.

## VI. CONCLUSION

In this paper, we conducted an experimental evaluation of the topological properties of two types of composition networks, i.e. dependency and interaction networks. Furthermore, syntactic and semantic Web services descriptions of the SAWSDL-TC collection have been used in order to extract the networks. Results show that dependency and interaction networks share the majority of their topological properties: the same component organization (a large one, a set of small ones and isolated nodes), the small-world property (each parameter or operation can be reach by a short path from any other), the disassortative degree correlation (poorly connected parameters or operations are attracted by richly connected ones), the existence of hubs and authorities (important parameters that are shared by several operations, or important operations that share their parameters with many others). From this study, it appears that the dependency and the interaction networks models exhibit very similar topological structure. There is however a significant difference. While the degree distribution of the dependency networks follow a power-law distribution, this is not the case for the interaction networks. Nevertheless, we still observe the presence of hubs and authorities. The distribution is inhomogeneous with heavy tail and therefore very far from the one observed in random networks.

In future work, we are planning to evaluate semantic networks based on different matching scheme. We plan also to investigate the community structure of the networks. In the longer run, we plan to take advantage of the results of this work in order to propose efficient algorithms for Web services composition search.

## REFERENCES

[1] S.-V. Hashemian and F. Mavaddat. "A Graph-Based Approach to Web Services Composition." Symposium on Applications and the Internet, Trento, Italy, 2005.

[2] J. Liu and L. Chao. "Web Services as a Graph and Its Application for Service." Discovery.International Conference on Grid and Cooperative Computing (GCC). Changsha, Hunan, China, 2006.

[3] J. Gekas, and M. Fasli. "Employing Graph Network Analysis for Web Service Composition." Agent Technologies and Web Engineering. D. C. e. Alkhatib G. I. and Rine, IGI Global, 2008.

[4] H. N. Talantikite, D. Aissani and N. Boudjlida. "Semantic annotations for web services discovery and composition." Computer Standards & Interfaces 31(6): 1108-1117, 2009.

[5] H. Kil, S. C. Oh, E. Elmacioglu, W. Nam and D. Lee. "Graph Theoretic Topological Analysis of Web Service Networks." World Wide Web Vol. 12, No. 3: page 321-343, 2009.

[6] S. C. Oh and D. Lee. "WSBen: A Web Services Discovery and Composition Benchmark Toolkit." International Journal of Web Services Research (JWSR) Vol.6, No.1: 1-19, 2009.

[7] G. Alonso, F. Casati, H. Kuno and V. Machiraju. Web Services. "Concepts, Architectures and Applications." Springer, 2004.

[8] T. Berners-Lee, J. Hendler and O. Lassila. "The semantic web." In Scientific American, May 2001.

[9] R. Akkiraju et al.. "Web Service Semantics - WSDL-S." 2005. Retrieved January, 2011, from http://www.w3.org/Submission/WSDL-S/.

[10] J. Farrell, and H. Lausen. "Semantic Annotations for WSDL and XML Schema." 2007. Retrieved January, 2011, from http://www.w3.org/TR/sawsdl/.

[11] D. Martin et al. "OWL-S: Semantic Markup for Web Services." 2004. Retrieved January, 2011, from http://www.w3.org/Submission/OWL-S/.

[12] C. Cherifi, V. Labatut, and J. F. Santucci, "On Flexible Web Services Composition Networks," in Digital Information and Communication Technology and Its Applications, 2011, vol. 166, pp. 45-59.

[13] M. Paolucci, T. Kawamura, T.R. Payne and K. Sycara. "Semantic Matching of Web Services Capabilities." International Semantic Web Conference on The Semantic Web (ICSW). Sardinia, Italy, Springer-Verlag: 333-347, 2002.

[14] M.-E.-J. Newman. "The structure and function of complex networks." SIAM Review **45**: SIAM Review, 2003.

[15] R. Albert, H. Jeong and A.-L. Barabási (1999). "The diameter of the world wide web." Nature **401**: 130.

[16] S. Boccaletti, V. Latora, Y. Moreno, M. Chavez and D.-U. Hwang. "Complex networks: Structure and dynamics." Physics Reports 424: 175—308, 2006.

[17] D. J. Watts and S. H. Strogatz. "Collective dynamics of small-world networks." Nature 393: 440—442, 1998.

[18] C. Cherifi, V. Labatut, and J. F. Santucci, "Web Services Dependency Networks Analysis." In International Conference of New Media and Interactivity (NMI 2010), 2010, pp. 115-120.

[19] C. Cherifi, V. Labatut, and J. F. Santucci, "Benefits of Semantics on Web Service Composition from a Complex Network Perspective." In International Conference on Networked Digital Technologies (NDT 2010), 2010, vol. 88, pp. 80-90.

[20] A. Hess, E. Johnston and N. Kushmerick. "ASSAM: A Tool for Semi-Automatically Annotating Semantic Web Services." International Semantic Web Conference (ISWC) Hiroshima, Japan, 2004.

[21] ICEBE'05. "IEEE International Conference on e-Business Engineering (ICEBE)." 2005. Retrieved January, 2011, from http://ieeexplore.ieee.org/xpl/mostRecentIssue.jsp?punumber=10403.

[22] InfoEther and B. Technologies. "SemWebCentral.org website." 2004. Retrieved January, 2011, from http://wwwprojects.semwebcentral.org/.

[23] U. Küster, B. König-Ries and A. Krug. "OPOSSum - An Online Portal to Collect and Share SWS Descriptions." International Conference on Semantic Computing (ICSC), 2008.

[24] C. Cherifi, Y. Rivierre, and J.-F.Santucci. "WS-NEXT, a Web Services Network Extractor Toolkit." In International Conference on Information Technology (ICIT'11). Amman, 2011.